\documentstyle[11pt,aaspp,epsf,rotate]{article}
\eqsecnum
\def\beq{\begin{equation}}
\def\eeq{\end{equation}}
\def\simge{\mathrel{%
   \rlap{\raise 0.511ex \hbox{$>$}}{\lower 0.511ex \hbox{$\sim$}}}}
\def\simle{\mathrel{
   \rlap{\raise 0.511ex \hbox{$<$}}{\lower 0.511ex \hbox{$\sim$}}}}

\begin{document}
\title{Evidence for Hot, Diffuse Gas in the Local Supercluster}
\author{Stephen P. Boughn}
\affil{Department of Astronomy, Haverford College, Haverford, PA  19041
sboughn@haverford.edu}

\begin{abstract}
The HEAO1 A2 full sky, $2-10~keV$ X-ray map was searched for emission
correlated with the plane of the local supercluster of galaxies.  After 
removing strong point and moderately extended sources (e.g. the core of the 
Virgo cluster), there remained a statistically signficant component
of ``diffuse'' X-rays in the plane of the supercluster.  Fitting this diffuse
component with a simple ``pillbox'' model of the local supercluster implies
a volume X-ray emissivity of $\varepsilon_x = 3.0 \pm 0.3 \times 10^{39}~
(R_{SC}/20~Mpc)^{-1}erg~ s^{-1}Mpc^{-3}$ where $R_{SC}$ is the radius of the 
supercluster and the error is photon counting noise only.  If one considers 
fluctuations in the X-ray background as an additional component of noise then
the significance of the detection is reduced to $2$ to $3 \sigma$.  This is 
consistent with fits of the model to data sets obtained by rotating the original
data.  The distribution of these rotated fits indicates that the detection is 
signficant at the $99\%$ confidence level.  If the source of the X-ray emission 
is Bremsstrahlung from a uniformly distributed plasma with temperature
$T_e$ then the implied electron number density is 
$N_e = 2.5 \times 10^{-6}~(R_{SC}/20~Mpc)^{-{1\over 2}}~(kT_e /10~keV)
^{-{1\over 4}}~cm^{-3}$.  This value is about an order of magnitude larger 
than the average baryon number density implied by nucleosynthesis and is 
consistent with a collapse factor of $10$.  A search for similar structure in
the COBE $53~GHz$ microwave background map yielded a marginal detection with
an amplitude of $\sim -17\pm 5~\mu K$ (statistical error only) which is 
consistent with the Sunyaev-Zel'dovich (SZ) effect expected from $10~keV$ gas.
This latter value is comparable to the amplitude of intrinsic large-scale
fluctuations in the microwave background and should be considered to
be a $1\sigma$ result at best.

\end{abstract}


\section{Introduction}

The largest known structures in the clustering heirarchy of galaxies are 
flattened distributions known as ``superclusters'' (SCs).  The local supercluster
(LSC), the supercluster of which the Galaxy is a member, was initially discussed
by de Vaucouleurs (1953).  Since then catalogues containing hundreds of SCs have
been compiled (\cite{bs84}; \cite{bb85}; \cite{ein97}).  
Although it seems likely that many, if not most, SCs are gravitationally 
bound structures, they are far from virialized.  Never-the-less, Small et al. 
(1998) have recently used virial type arguments to place a lower limit on the
mass of the Corona Borealis SC.  The fact that SCs are only marginally
over-dense, i.e. $\delta \rho / \rho \sim 10$, further complicates studies of
their structures.  While the study of SCs is in its infancy, these objects 
promise to provide important information about the formation of large-scale
structure in the Universe.

If little is known about the dynamics of SCs, even less know about the 
intra-supercluster (ISC) medium.  Assuming SCs have baryon-to-light ratios 
comparable to rich clusters of galaxies, then one expects a substantial 
amount of ISC gas.  Furthermore, the virial temperatures of SCs are $\sim 10^8~K$ 
so it would not be surprising to find ``hard'' ($> 1~keV$) 
X-ray  emission from the ISC medium as is the case for the intergalactic medium 
in rich clusters of galaxies.  Most models of structure formation predict the
presence of a hot ISC, whether it is primoridal, created from winds from an
early population of stars, or tidally stripped from merging structures 
(\cite{mb98} and references cited therein).  Estimated
temperatures range from $10^6~K$ to $10^8~K$ (\cite{kk91}; \cite{rp92};
\cite{me94}; \cite{an96}).
It is straightforward to show that the cooling time for such gas at
the expected densities ($< 10^{-3}~cm^{-3}$) is much longer than a Hubble
time (\cite{rp92}) so it would remain hot today.

There have been several searches for diffuse X-ray emission from SCs.  Although
Murray et al. (1978) claimed that UHURU data showed evidence for ISC emission,
subsequent observations and analyses have not supported this claim 
(\cite{pra79}; \cite{prb88}; \cite{per90}).
Using ROSAT PSPC data Bardelli et al. (1996) reported 
the detection of diffuse emission 
in the region between two clusters in the Shapley supercluster.  On the other
hand, Day et al. (1991) and Molnar \& Birkinshaw (1998) have placed relatively 
strong upper limits on the $2-10~keV$ diffuse emission in this supercluster.  
While the current upper limits are interesting,  they still leave room for a 
substantial amount of hot ISC matter and more sensitive searches are underway.

Particularly intriguing is the result from the cross-correlation analysis of
the ROSAT All-Sky Survey with the Abell catalog.  Soltan et al. (1996) found
that Abell clusters seemed to be associated with diffuse X-ray emission with 
an extent of $\sim 20 Mpc$.  It is tempting to associate this emssion with 
hot ISC gas; although, there is as yet no direct evidence for this.

Hot ISC gas can also leave its signature on the Cosmic Microwave Background 
(CMB) via the Sunyaev-Ze'dovich (SZ) effect.  Hogan (1992) suggested that
the SZ effect in superclusters might account for the fluctuations in the 
CMB observed by the COBE satellite; however, subsequent analyses have shown that
this is not the case (\cite{bj93}; \cite{ben93}).
Molnar and Birkinshaw (1998)
used the lack of an SZ effect in the COBE DMR data to place an additional 
constraint on the ISC gas in the Shapley supercluster while  
Banday et al. (1996) found no evidence for the SZ effect in superclusters 
using a statistical cross-correlation analysis.  Because of the large 
intrinsic fluctuations in the CMB, searches for the SZ effect in SCs 
will undoubtedly result in upper limits until the next generation of CMB 
satellites provide adequate frequency coverage to resolve the two effects.

Searches for ISC gas in the local supercluster (LSC) have the advantage of 
much higher integrated signal, however, are frustrated by the presence of 
other large-scale structure in the X-ray sky.  Shafer (1983) and Boldt (1987) 
reported evidence for large-scale structure in the HEAO 1 A2 $2-10~keV$ data 
which is roughly consistent with either the Compton-Getting dipole expected
from the Earth's motion with respect to the CMB or general emission from the 
direction of the center of the LSC.  Subsequent analyses (\cite{sf83})
demonstrated that Compton-Getting dipole adequately accounts for this
large-scale structure; however, the direction of the dipole was determined to
be in a somewhat different direction than that 
of the CMB dipole.  Jahoda and Mushotzky
(1989) found evidence for enhanced $2-10~keV$ emission from the direction of 
the Great Attractor which is in the same general direction as the
Compton-Getting dipole.  In addition, Jahoda (1993) has found evidence for
high lattitude $2-10~keV$ emission from the Galaxy as well as emission 
associated with the Superglactic plane (the plane of the LSC), the latter of 
which is particularly relevant to the analysis presented in the this paper.

In the following sections we present evidence that there is enhanced emission in
the Supergalactic plane and that this emission is distinct from the other 
large-scale structures indicated above.  To the extent that this is true, we will
argue that the emission is diffuse and not associated with galaxies or other 
compact X-ray sources in the plane.  It is, of course, possible that the
emission is due to still larger scale structure of the type discussed by 
Miyaji \& Boldt (1990) and Lahav, Piran, \& Treyer (1997) or even to the 
chance allignment of fluctuations in the X-ray background (XRB); although, we 
argue that the latter is unlikely.  These possibilites will be discussed in 
more detail below.

\section{HEAO1 A2  ${\mathbf 2 - 10~keV}$ X-ray Map}

The HEAO1 A2 experiment measured the surface brightness of the X-ray
background in the $0.1 - 60~keV$ band (\cite{bol87}).  The present data set was
constructed from the output of two medium energy detectors (MED) with different
fields of view ($3^\circ \times 3^\circ$ and $3^\circ \times 1.5^\circ$) and 
two high energy detectors (HED3) with
these same fields of view.  The data were collected during the six month period
beginning on day 322 of 1977.  Counts from the four detectors were combined and 
binned in 24,576  $1.3^\circ \times 1.3^\circ$ pixels in an equatorial 
quadrilateralized spherical cube projection on the sky (\cite{ws92}).
The combined map has a spectral resolution of approximately 
$2 - 10~keV$ (\cite{jm89}) and is shown in Galactic 
coordinates in Figure 1.

\begin{figure}[ht]
\centerline{\rotate[r]{\vbox{\epsfysize=17cm\epsfbox{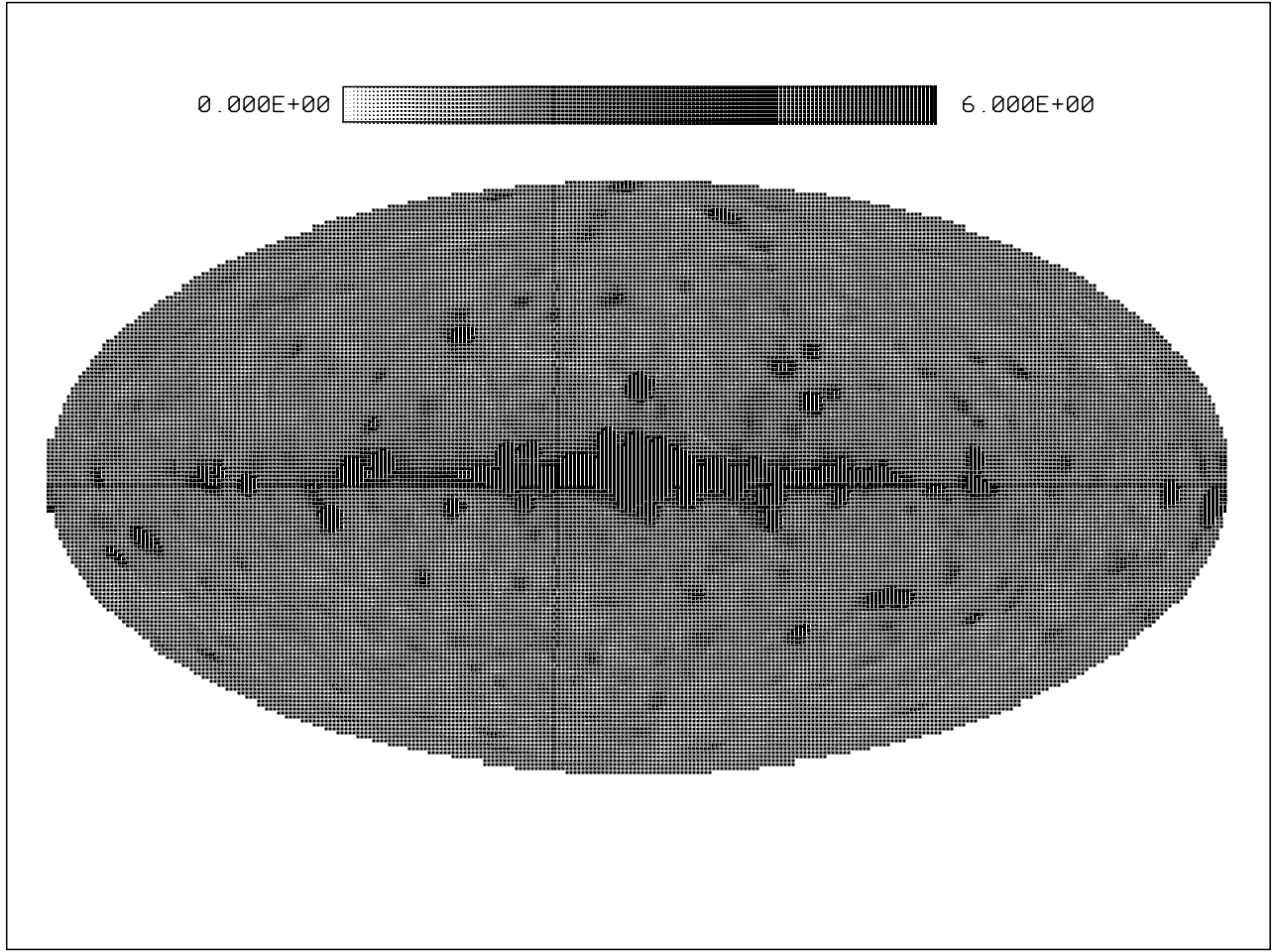}}}}
\caption[]
{
$~~2-10~keV$ HEAO I A2 map in Galactic coordinates (Jahoda \& Mushotzky 1989).
The units are $TOT~cts/sec/4.5~deg^2$ (see text).
}
\end{figure}

The effective point spread function (PSF) of the map was determined by averaging
the PSFs of 75 HEAO1 point sources (\cite{pic82}).  The composite PSF
is well fitted by a gaussian with a full width, half maximum of $2.96^\circ$.  
This value will be important in estimating the effective noise in the maximum
liklihood fits of \S3.4.  Because of the pixelization, the PSF varies
somewhat with location on the sky; however, this has little effect on noise
estimates so we assume a constant PSF.

In order to remove the effects of both points source and large-scale
structure, more than half the map was flagged and not included in the fits
described in \S3.  The dominant feature in the HEAO map is the Galaxy (see
Figure 1) so all data within 
$20^\circ$ of the Galactic plane and, in addition, within $30^\circ$ of the 
Galactic center were cut from the map.  In addition, $10^\circ$ diameter regions
 around $90$ discrete X-ray sources with $2 - 10~keV$ fluxes larger than 
$3 \times 10^{-11} erg~s^{-1} cm^{-2}$ (\cite{pic82}) were removed.
The resulting ``cleaned'' map covered about 50\% the sky.  In order to identify
additional point sources, the map itself was searched for ``sources'' that 
exceeded the nearby background by a specified amount.  This was accomplished by
first averaging each pixel with its $8$ neighbors (note that the
quadrilaterized cube format lays out the pixels on an approximately square
array) and then comparing this value with the median value of the next nearest
$16$ pixels.  Pixels within the flagged Galactic region are ignored.  If the
average flux associated with a given pixel exceeds the median flux of the
background, then all 25 pixels are flagged and removed from further
consideration.  These flagged regions correspond approximately to a $6.5^\circ 
\times 6.5^\circ$ patch.  This proceedure is not done iteratively, i.e. each 
comparison is made on the basis of the original map with only the Galaxy 
flagged.  Cuts were made at several levels corresponding to average fluxes of 
from $3$ to $7\times 10^{-12}~erg~s^{-1}cm^{-2}$.  The most extreme cut 
corresponds to an equivalent point source flux of 
$3\times 10^{-11}~erg~s^{-1}cm^{-2}$ and results in
$75\%$ of the sky being flagged.  At this level, most of the ``sources'' 
cut are either noise fluctuations or fluctuations in the X-ray background.
In any case, the results of \S3.5 are largely insensitive to these cuts.

Only one large-scale correction to the map, the Compton-Getting dipole, was
made apriori.  If the dipole moment of the
cosmic microwave background is a kinematic effect, as it has been widely
interpreted (\cite{ben96}), then the X-ray background should possess
a similar dipole structure (Compton-Getting effect) with an amplitude of
$4.3 \times 10^{-3}$.  As discuseed in \S1, evidence for this structure is
found in the HEAO map (\cite{sha83}; \cite{sf83}; \cite{lpt97}).
The cleaned map was corrected for this effect; however, the
results of \S3.5 are the same even if the amplitude and direction of the dipole
are fit from the data.

The presence of three other sources of large-scale structure in the X-ray 
map has been noted.  A linear time drift in detector sensivity (\cite{jah93}) 
results in effective large-scale structure of known form but unknown amplitude.
In addition, the $2 - 10~keV$ background shows evidence of high latitude 
Galactic emission as well as emission associated with the Supergalactic plane 
(\cite{jah93}).  Models for all three of these contributions are simultaneously 
fit to the data as is discussed in \S3.4.  

Because of the ecliptic longitude scan pattern of the HEAO satellite, sky
coverage and, therefore, photon shot noise are not uniform.  However, the
mean variance of the cleaned, corrected map, 
$2.0 \times 10^{-2} (TOT~cts/sec)^2$, 
is considerably larger than the mean variance of photon shot noise, 
$0.67 \times 10^{-2} (TOT~cts/sec)^2$, where 
$1~TOT~ct/sec \approx 2.1 \times 10^{-11} 
erg~s^{-1} cm^{-2}$ (\cite{ajw94}).  This implies that 
the X-ray map is dominated by ``real'' structure (not photon shot noise).  
For this reason, we chose to weight the pixels equally in all subsequent 
analyses.

\section{Modelling the Large-Scale Structure}
\subsection{Instrument Drift}
At least one of the A-2 detectors changed sensivity by $\sim 1\%$ in the six 
month interval of the current data set (\cite{jah93}).  Because of the ecliptic
scan pattern of the HEAO satellite, this results in a large-scale pattern
in the sky which varies with ecliptic longitude with a period of $180^\circ$.
If the drift is assumed to be linear, the form of the resulting large-scale 
structure in the map is completely determined.  A linear drift of unkown 
amplitude is taken into account by constructing a sky map with the appropriate 
structure and then fitting for the amplitude simultaneously with the other 
parameters.  We investigated the possibility of non-linear drift by considering
quadratic and cubic terms as well; however, the results of \S3.5 were insensitive 
to this refinement.
\subsection{The Galaxy}
The X-ray background has a diffuse (or unresolved) Galactic component which 
varies strongly with Galactic latitude (\cite{iwa82}).  This emission is
still significant at high Galactic latitude ($b_{II}>20^\circ$) and extrapolates
to $\sim 1\%$ at the Galactic poles.  We modeled this emission in two ways.  The
first model consisted of a linear combination of a secant law Galaxy with 
the Haslam $408~GHz$ full sky map (\cite{has82}). 
The latter was included to take into 
account X-rays generated by inverse Compton scattering of CMB photons 
from high energy electrons in the Galactic halo, 
the source of much of the synchtron emission in the Haslam map.  
We find only marginal evidence for such emission (\S3.5).  As an alternative 
Galaxy model we also considered the two disk, exponentially truncated model
of Iwan et al. (1982).  The analysis of \S3.5 shows signficant X-ray emission 
corrlelated with either of these models.
\subsection{The Local Supercluster}
The level of emission from the plane of the LSC reported in this paper is 
barely above the noise ($S/N\sim 3$).  Therefore, detailed models of
LSC emission are not particularly useful.  We chose a simple ``pillbox'' 
model, i.e uniform X-ray emissivity, $\varepsilon_x$,  
within a circular disk of radius $R_{SC}$
and height (thickness), $H_{SC}$.  The X-ray intensity in a particular 
direction is then proportional to the pathlength, $L$, through the LSC disk, 
i.e., $I_x = \varepsilon_x L/4\pi$.  The nominal location of the LSC disk was
chosen to be in the Supergalactic plane (\cite{tul82}) with a nominal center 
in the direction
of the Virgo cluster (M87); however, these positions were allowed to vary from
their fiducial locations.  The radial position of the Galaxy was assumed to be 
$ 0.8~R_{SC}$ from the center of the LSC, i.e., near the edge.
The value of $R_{SC}$ is left as a scale 
parameter in the final results; however, the dependence on the diameter of the 
LSC was investigated by varying the radial position of the Galaxy within the
LSC disk (see \S4.5).  The thickness to diameter ratio, $H_{SC}/2R_{SC}$, was 
also varied; however, we
considered the nominal value to be $1/8$ which is consistent with the
distribution of galaxies in the LSC (\cite{tul82}).  In any case, as is true for
the Galaxy model, the results of \S3.5 are not overly sensitive to these model
parameters.

\subsection{Analysis}
Combining the three above models with a uniform X-ray background we arrive at
the following 5 parameter expression for the ``diffuse'' X-ray intensity, $X_i$, 
in the $i^{th}$ sky pixel,
\begin{equation}
X_i = a_1 + a_2\times P_i + a_3\times T_i + a_4\times S_i + a_5\times H_i
\end{equation}
where the first term represents the intensity of the uniform X-ray background, 
$P_i$ 
represents the intensity due to the LSC ``pillbox'' normalized to a pathlength 
of $1~R_{SC}$, $T_i$ is the pattern on the sky caused by a linear drift in
detector sensitivity, $S_i$ is emission proportional to the cosecant of Galactic
latitude normalized to the Galactic pole, $H_i$ is the antenna temperature
(in $^\circ K$) of the $408~MHz$ Haslam map, and $a_k$ are the 5 free
parameters.  $P_i$, $S_i$, and $H_i$ are 
convolved with the PSF of the observed map. Since, as discussed above, 
each pixel was weighted equally, the
least squares fit to the 5 parameters is obtained by minimizing an effective
$\chi^2$, 
\begin{equation}
\chi^2 =  \sum_i (x_i - \sum_m a_m X_{m,i})^2
\end{equation}
where $x_i$ is the observed X-ray intensity in the $i^{th}$ pixel, $X_{m,i}$ 
is the $i^{th}$ pixel of the $m^{th}$ parameter map (i.e., $X_{1,i}=1$, 
$X_{2,i}=P_i$, $X_{3,i}=T_i$, $X_{4,i}=S_i$, and $X_{5,i}=H_i$), and $a_m$ is
the $m^{th}$ fit parameter.  The sum, $\sum_i$, is over all unflagged
pixels in the cleaned HEAO map.  

The errors in the fit due to uncorrelated shot
noise are easily computed by (see for example Press et al. 1986)
\begin{equation}
{\sigma_k}^2 = \sum_{m,n} C_{k,m}C_{k,n} \sum_i {\sigma_i}^2 X_{m,i}X_{n,i}
\end{equation}
where $1 \leq k,m,n \leq 5$ indicate the 5 fit parameters and $\sigma_i$
indicates the photon shot noise in the $i^{th}$ pixel.  $C_{m,n}$ is the
inverse of the matrix $A_{m,n}$ which is defined as 
$A_{m,n}=\sum_{i} X_{m,i} X_{n,i}$.  

The intrinsic fluctuations in the X-ray background (XRB) can be thought of as an
additional source of noise.  The errors in the fit parameters induced
by this noise are more problematic since 
XRB fluctuations exhibit pixel to pixel correlations due to the finite PSF
of the detectors and the clustering of X-ray sources.  In addition, these
fluctuations may not be uniform on the sky.  If 
$R_{i,i^\prime} = \langle \delta I_{x,i} \delta I_{x,{i^\prime}} \rangle$ 
represents the auto-correlation function (ACF) of the intensity of the
XRB then it is straightforward to show that the corresponding errors 
in the fit parameters are given by
\begin{equation}
{\sigma^\prime _k}^2 = \sum_{m,n} C_{k,m}C_{k,n} 
\sum_{i,i^\prime} R_{i,i^\prime} X_{m,i}X_{n,i^\prime}
\end{equation}

As an estimate of these errors we will, in the analysis to follow, assume that
the fluctuations are uniform.   Figure 2 is the ACF of the X-ray map (map$\#2$
 - see \S3.5) corrected for large-scale structure (i.e., a uniform
X-ray background, the Compton-Getting dipole, Galaxy and LSC emission, and 
instrumental drift).  The point at $\theta = 0$ has been corrected for 
photon shot noise.  If correlated structure on small angular scales is
entirely due to a gaussian PSF, i.e.,
$PSF \propto e^{-\theta ^2 / 2\sigma_p ^2}$, it is straightforward 
to show that the ACF has the form, 
$R_{i,i^\prime}=R_\circ e^{-\theta_{i,i^\prime} ^2 /4\sigma_p ^2}$ where 
$\theta_{i,i^\prime}$ is the angle between the $i^{th}$ and the $i^{\prime th}$
pixels.  The dashed curve in Figure 2 is this functional form.  
Note that $\sigma_p$ is not a fit
parameter but is determined by the profiles of point sources (see \S2).  While 
this curve is a good fit to the data for $\theta < 4^\circ$, the data
for $4^\circ \simle \theta \simle 10^\circ$ lie somewhat above the curve. 
Whether or not this is due to clustering of the X-ray background or to an
improperly modeled PSF is not clear; however, it seems unlikely that the PSF
extends out to $10^\circ$.  In any case, we have also modelled the ACF as the
sum of an exponential and a term proportional to $(\theta ^2 + \theta_o ^2)^{-1}$
to account for the possibility of large-scale clustering.  The solid curve in
Figure 2 represents this four parameter fit.

\begin{figure}[ht]
\centerline{\rotate[r]{\vbox{\epsfysize=15cm\epsfbox{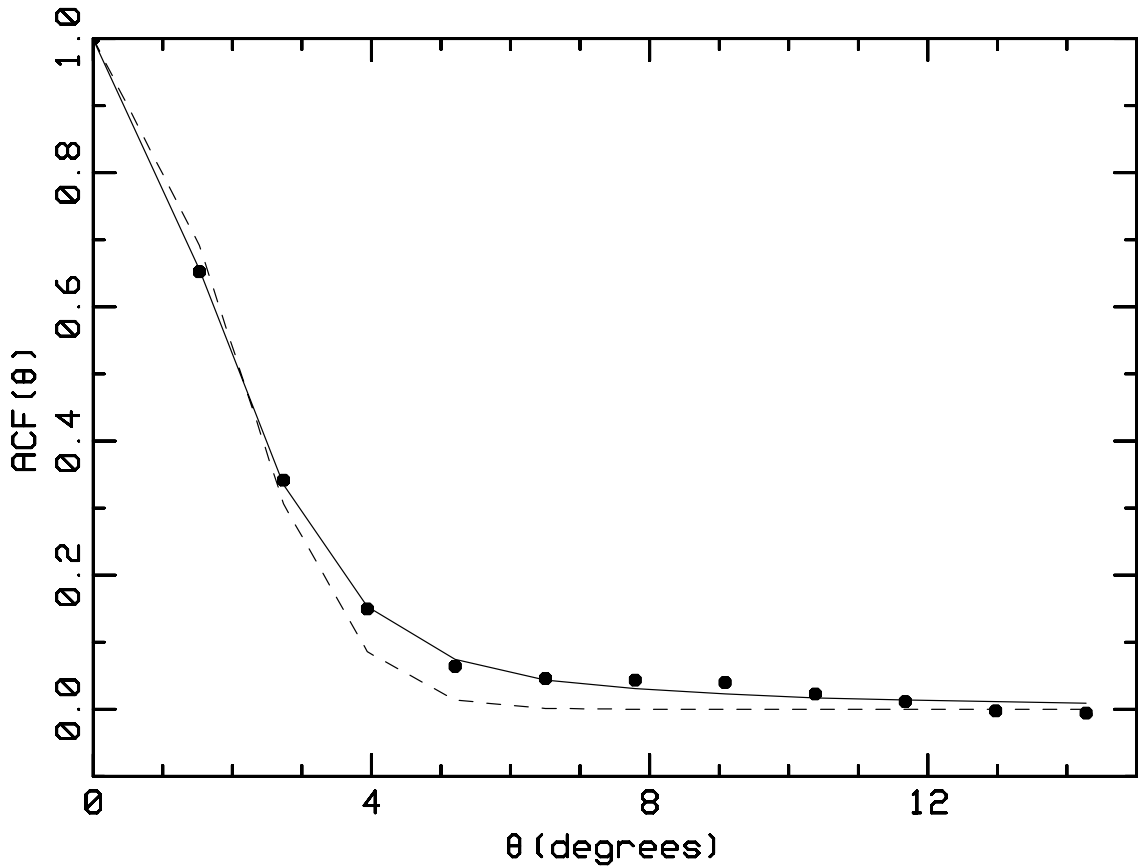}}}}
\caption[]
{
The autocorrelation function (normalized to unity) of Map$\#2$ (see \S3.5) 
corrected for large-scale structure.  The dashed curve is that expected for
the Gaussian PSF of the map.  The solid curve is a fit to the data to
account for the tail of the ACF (see text).
}
\end{figure}

Since the photon shot noise and sky fluctuations are uncorrelated, the
two sources of errors should be added in quadrature, i.e.,
$\sigma_{tot} ^2 = \sigma ^2 + \sigma^{\prime 2}$.

\subsection{Results}
	The results of the analysis described in \S3.4 for two different 
cleaning windows of the map are given in Table 1.  Map$\#1$ is minimally 
windowed, i.e. the Galaxy pixels are flagged as well as the regions
around the Piccinotti sources (see \S2).  In addition, 20  ``sources'',
i.e., isolated regions of high flux, were flagged (see \S2); however, these 
comprised only 0.21 sr, i.e. 1.6\% of full sky.   This cleaned map 
contains 11637 pixels which corresponds to a 47\% sky coverage.  Map$\#2$ was
subjected to more aggressive ``source'' cleaning and contains 8153 pixels, i.e.
33\% sky coverage.

The three errors listed in Table 1 are for photon shot noise only and for shot noise 
plus the estimated noise for fluctuations in the XRB according to the
two models of Figure 2 (see \S3.4).
For shot noise alone, $\chi_\nu^2$ is quite large which is simply an indication of 
the importance of fluctuations in the XRB.  When these fluctuations are included
$\chi_\nu^2$ drops considerably.  The fact that in the latter case $\chi_\nu^2 \sim 1$
is not particularly signficant since fluctuations in the XRB were esimated from
excess noise in the map.  The
effective number of degrees of freedom is considerably less (by about a
factor of 10) than the number of 
pixels since the largest component of the noise is correlated.

It is clear from Table 1 that the parameters from the two fits are 
consistent with each other and indicate (with the possible expection of $a_5$)
that all parameters are significantly different from zero.  From Table 2 we
see that fit parameters are not strongly correlated with each other.  Of 
particular significance to this paper is that the flux associated with the
supercluster is positive by $\sim 3 \sigma$ even when the fluctuations in the 
XRB are taken into account.  A test for the statistical significance for this
result is discussed in \S4.4.  Taken at face value, the fits in Table 1
indicate that diffuse X-ray emission from the local supercluster has been
detected.
However, a great many checks must be made on robustness of this result: e.g.,
whether or not the signal comes from a few strong point sources located near the
supergalactic plane; whether or not the signal is due X-ray emitting galaxies
distributed in the LSC; whether or not the signal arises from a chance
allignment of fluctuations in the X-ray background; whether or not the result 
is sensitive to the pillbox model parameters; etc.  The next section 
contains a detailed discussion of these issues.

\begin{table}[h]
\begin{center}
\begin{tabular}{c|rrrrccc|rrrr}
\multicolumn{5}{l}{Map$\#1$} & \multicolumn{2}{l}{    } &
\multicolumn{5}{l}{Map$\#2$} \\
\multicolumn{1}{c}{$k$} & \multicolumn{1}{c}{$a_k$} &
\multicolumn{1}{c}{$\sigma_a$} & \multicolumn{1}{c}{$\sigma_b$} &
\multicolumn{1}{c}{$\sigma_c$} & \multicolumn{2}{l}{    } &
\multicolumn{1}{c}{$k$} & \multicolumn{1}{c}{$a_k$} &
\multicolumn{1}{c}{$\sigma_a$} & \multicolumn{1}{c}{$\sigma_b$} &
\multicolumn{1}{c}{$\sigma_c$} 
\\ \cline{1-5} \cline{8-12}
1 & 328.17 & 0.36 & 1.30 & 1.49 & & & 1 & 329.06 & 0.42 & 1.49 & 1.73 \\
2 &   3.38 & 0.29 & 1.06 & 1.18 & & & 2 &   3.16 & 0.34 & 1.21 & 1.37 \\
3 &   6.38 & 0.29 & 0.99 & 1.06 & & & 3 &   6.26 & 0.34 & 1.15 & 1.30 \\
4 &   3.23 & 0.18 & 0.64 & 0.70 & & & 4 &   2.25 & 0.21 & 0.74 & 0.84 \\
5 &   0.10 & 0.01 & 0.05 & 0.05 & & & 5 &   0.06 & 0.02 & 0.05 & 0.06 
\\ \cline{1-5} \cline{8-12}
$\chi_\nu^2$ &  & 3.57 & 0.99 & 0.99 & & & $\chi_\nu^2$ &  & 2.96 & 1.00 & 1.00
\end{tabular}
\end{center}
\caption{Fit parameters for Maps$\#1$ and $\#2$.  The units are 
$0.01~TOT~cts/sec/4.5~deg^2 \approx 1.54 \times 10^{-10}~
erg~s^{-1}cm^{-2}sr^{-1}$.
The parameter $a_1$ corresponds to the intensity of the XRB, $a_2$ is emission
from the LSC, $a_3$ is detector drift, $a_4$ is Galactic secant law emission,
and $a_5$ is emission proportional to the Haslam map (in units of 
$0.01~TOT~cts/sec/4.5~deg^2/K$).  $\sigma_a$ is the error in the fit for 
photon counting
noise only, $\sigma_b$ includes fluctuations in the XRB modeled as the dashed
curve in Figure 2, and $\sigma_c$ includes fluctuations modeled as the solid
curve of Figure 2.}
\end{table}

\begin{table}[h]
\begin{center}
\begin{tabular}{c|rrrrr}
\multicolumn{1}{c}{ } & \multicolumn{1}{c}{$a_1$} & \multicolumn{1}{c}{$a_2$} &
\multicolumn{1}{c}{$a_3$} & \multicolumn{1}{c}{$a_4$} &
\multicolumn{1}{c}{$a_5$} \\ \cline{2-6}
$a_1$ &  1.0 & -0.4 & -0.1 & -0.5 & -0.5 \\
$a_2$ & -0.4 &  1.0 &  0.0 &  0.2 &  0.0 \\
$a_3$ & -0.1 &  0.0 &  1.0 & -0.3 &  0.0 \\
$a_4$ & -0.5 &  0.2 & -0.3 &  1.0 & -0.3 \\
$a_5$ & -0.5 &  0.0 &  0.0 & -0.3 &  1.0
\end{tabular}
\end{center}
\caption
{Correlation coefficients for the fit parameters for Map$\#1$ in 
Table 1.  The fit parameters, $a_k$, are defined as in Table 1.
The coefficients for Map$\#2$ are similar.}
\end{table}

\section{Systematics}
\subsection{The Galaxy}
Even though the plane of the Galaxy, a strong source of X-ray emission, was
removed from the map (see \S2), high Galactic latitude emission is large 
enough to be a potential source of error.  
However, the planes of the Galaxy and the LSC are nearly perpendicular and the 
center of the LSC is nearly at the Galactic pole.  It is these fortuitous 
circumstances that result in nearly uncorrelated Galaxy and LSC fit 
parameters (see Table 2).  As a consequence, one expects that the fit to LSC
emission will be nearly independent of Galactic emission.  
As an alternative Galaxy model we considered the two disk, exponentially 
truncated model of Iwan et al. (1982).  The $\chi^2$ for the fits with this
model were slightly worse while the LSC fit parameters were essentially 
unchanged ($\delta a_2 \simle 3\%$).  Even if the Galaxy model is left out of 
the fit entirely, the resulting LSC emission changes by only $\sim 25\%$; 
however, the $\chi^2$ is signficantly worse.  Therefore, we consider it unlikely
that the supercluster emission found above is due to Galaxy contamination.

\subsection{Compton-Getting Dipole}
As discussed in \S2 , the X-ray map was corrected for the Earth's motion 
relative to the average rest frame of the distant Universe as defined by the
CMB dipole.  This is justified since the bulk of
the X-ray background arises from sources at high redshift ($z\simge 1$).  The 
amplitude of the X-ray dipole is easily computed from the CMB dipole and the
spectrum of the X-ray background, i.e., $\delta I_x / I_x = 4.3 \times 10^{-3}$
which is $\sim 3.4$ times larger than the CMB dipole.  In addition, it is 
likely that X-rays trace the asymmetric mass distribution 
which created our peculiar velocity, and one would expect an 
intrinsic dipole moment in the XRB which is more or less alligned with the
Compton-Getting dipole.  However, it seems unlikely that the former 
will be as large
as the latter.  From unified models of the XRB (e.g. \cite{com95}),
less than $0.2 \%$ of the XRB background arises from sources 
within $50~Mpc$ and with fluxes $< 3 \times 10^{-11} erg s^{-1} cm^{-2}$.
Lahav, Piran, and Treyer (1997) demonstrated that the large-scale structure
dipole can be comparable to the Compton-Getting dipole if one assumes, contrary
to observations, unevolving X-ray luminosity and no upper limit on source flux.
It is unlikely that their calculation is relevant to our analysis;
never-the-less, we considered the possibility that the dipole correction made 
to the data is in error.  If a dipole term is included in the fit, it is also 
not strongly correlated with the ``pillbox'' term and the resulting LSC 
emission changes by only $\sim 15\%$ for Map$\#1$ and by $\sim 5\%$ 
for Map$\#2$.  The $\chi^2$'s were only marginally better if
these terms are included.  It is interesting that the resulting fit dipoles for 
the two maps agree within errors with that predicted for the Compton-Getting 
dipole.  For Map$\#2$, the fit dipole amplitude is
$\delta I_x/ I_x = 5.3 \pm 1.8 \times 10^{-3}$ while the direction is $\sim
19^\circ$ from the CMB dipole which is well within the directional error.
For Map$\#1$ these values are $\delta I_x/ I_x = 2.7 \pm 1.8 \times 10^{-3}$ 
and $\sim 35^\circ$ also consistent with the Compton-Getting dipole.
We find the results of the dipole fits signficant in that they indicate 
that large-scale structure can be detected at this level.
If the dipole term is excluded altogether from the
fit, the values of LSC emission for the two maps increase by about $30\%$.

\subsection{Timedrift}
The timedrift fit parameter in Table 1 corresponds to somewhat less than
$2\%$ of the XRB during the six month interval of the observations and,
therefore, is also of potential concern.  However,
from Table 2 it is evident that the this fit parameter is essentially
uncorrelated with the LSC parameter indicating that our estimate of 
LSC emission is also not very sensitive to drift of detector
sensitivity.  In the case that the linear timedrift model is removed from the
fit, the LSC parameter changes by $< 5\%$ while the $\chi^2$'s are
significantly worse.  In order to test the limitations of the linear timedrift 
model we included cubic and quadratic terms as well.  The solutions
and $\chi^2$'s were essentially unchanged.  We conclude that detector drift 
cannot account for the observed emission from the LSC.

The Galaxy, the dipole, and the effects of timedrift are all more or
less orthogonal to the LSC.  In fact, if we remove all three of these items 
from the analysis the fit to LSC emission changes by only $4\%$ for Map$\#2$
and by $20\%$ for Map$\#1$.

\subsection{The XRB}
It is now generally accepted that the X-ray background is composed of discrete
sources (\cite{has93}; \cite{geo97}).  Source confusion
due to the finite resolution of the map along with inherent clustering of 
sources result in fluctuations in the XRB with an angular scale of about 
$3^\circ$ and an amplitude of about $3\%$ of the background.  On the other hand,
the implied LSC emission is about $1\%$ of the XRB so it is important to 
consider whether or not chance allignments of these fluctuation could account 
for it.

To the extent that fluctuations in the XRB can be modelled as in \S3.5,
the above analysis indicates
that the fit to LSC emission is signficant at the $\sim 3\sigma$ level.
As an independent check we repeated the fits for
5000 pillbox models with a uniform distibution of orientations in the sky.
Because of the extensive windowing of the Galactic plane, models lying within
$30^\circ$ of the Galactic plane were disregarded.  Also because of the
``detected'' LSC emission, models lying within $30^\circ$
of the LSC plane were also excluded.  For Map$\#1$, the amplitude of LSC emission
in Table 1 exceeds those of $97\%$ of the trials while $93\%$ of the $\chi^2$s
of the trial fits exceed the value in the table.  Only $2.6\%$ of the trials 
have LSC emission greater and $\chi^2$s less than those listed in Table 1.  The 
results for Map$\#2$ are even stronger.  The LSC emission exceeds the fits of
$99\%$ of the trials while the $\chi^2$ is exceeded by $97\%$ of the $\chi^2$s
of the trials.  Only $0.7\%$ of the trials have LSC emission greater and
$\chi^2$s less that the values listed in Table 1.  If one includes the models
that lie within $30^\circ$ of the plane of the LSC, the results for Map$\#2$
are essentially the same while for Map$\#1$ $4.6\%$ of the trials have LSC
emission greater and $\chi^2$s less that those listed in Table 1.
Since Map$\#1$ has significantly more ``hot spots'' than the more heavily 
windowed Map$\#2$, it is perhaps not surprising that the former is more 
susceptible to chance allignments of these ``sources'' resulting in
spuriously large amplitudes of emission accociated with several pillbox models.
We conclude that at the $\sim 99\%$ confidence level
fluctuations in the XRB are not responsible 
for the observed signal.  This corresonds to a $2\sigma$ to $3\sigma$ effect 
and is roughly consistent with the estimates of the errors indicated in Table 1.

\subsection{Pillbox Model}
The pillbox model for the LSC is highly simplistic; however,
more refined models seem umwarranted considering the relatively low signal 
to noise ($\simle 3\sigma$) indicated in Table 1.  Never-the-less, several 
important checks of the model can be made.  The checks illustrated in the
figures below are for the more extensively windowed Map$\#2$; however,
the conclusions are essentially the same for both maps.

Figure 3 is a plot of the LSC emission fit for a series of pillbox models 
lying in the supergalactic plane but with centers rotated by angles
in steps of $30^\circ$ from the nominal center of the LSC in Virgo.  
It is clear that the largest emission
signals occur when the model center is near the nominal direction.  In addition,
the $\chi^2$s of the fits are signficantly worse for models rotated by angles
$\geq 60^\circ$.  The lowest $\chi^2$ occurs (for both windowed maps) at an
angle of $-23^\circ$ for which the LSC emission is somewhat larger; however, 
the differences from the Virgo centered model are not significant.

\begin{figure}[ht]
\centerline{\rotate[r]{\vbox{\epsfysize=17cm\epsfbox{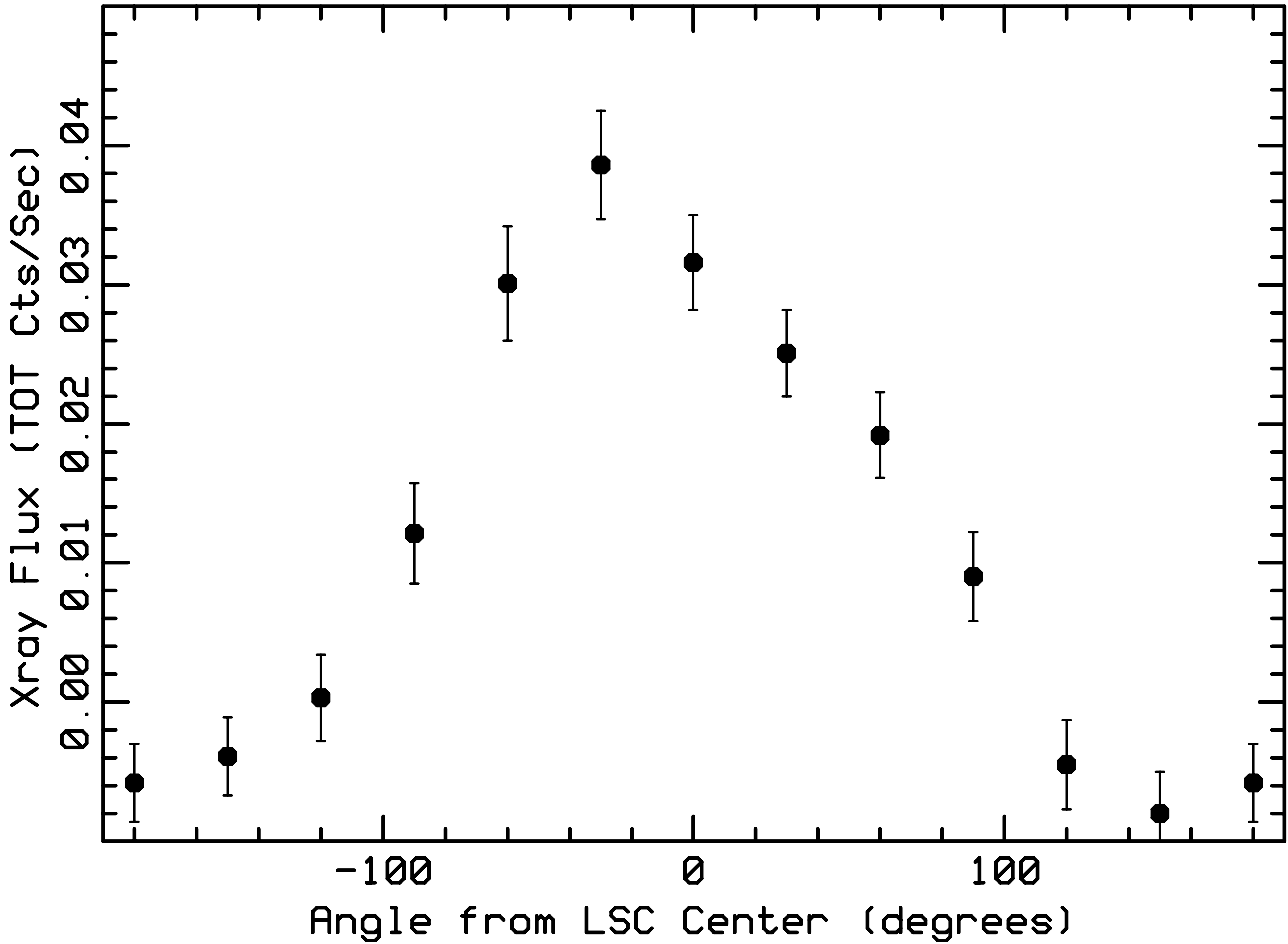}}}}
\caption[]
{
LSC emission fit to Map$\#2$ for pillbox models lying in the supergalactic plane
with centers rotated by angles in steps of $30^\circ$ from the nominal center 
of the LSC in Virgo.  The error bars are statistical only and are highly 
correlated.
}
\end{figure}

The dependence of the results on the vertical position of the Galaxy within
the disk of the LSC is illustrated in Figure 4.  The vertical position is
expressed in terms of the disk thickness, $H_{SC}$, so $\pm 0.5$ represents the
top and bottom of the disk.  Again the nominal central position yields the
largest LSC emission with the $\chi^2$s of the fits increasing signficantly
at the extremes.

\begin{figure}[ht]
\centerline{\rotate[r]{\vbox{\epsfysize=17cm\epsfbox{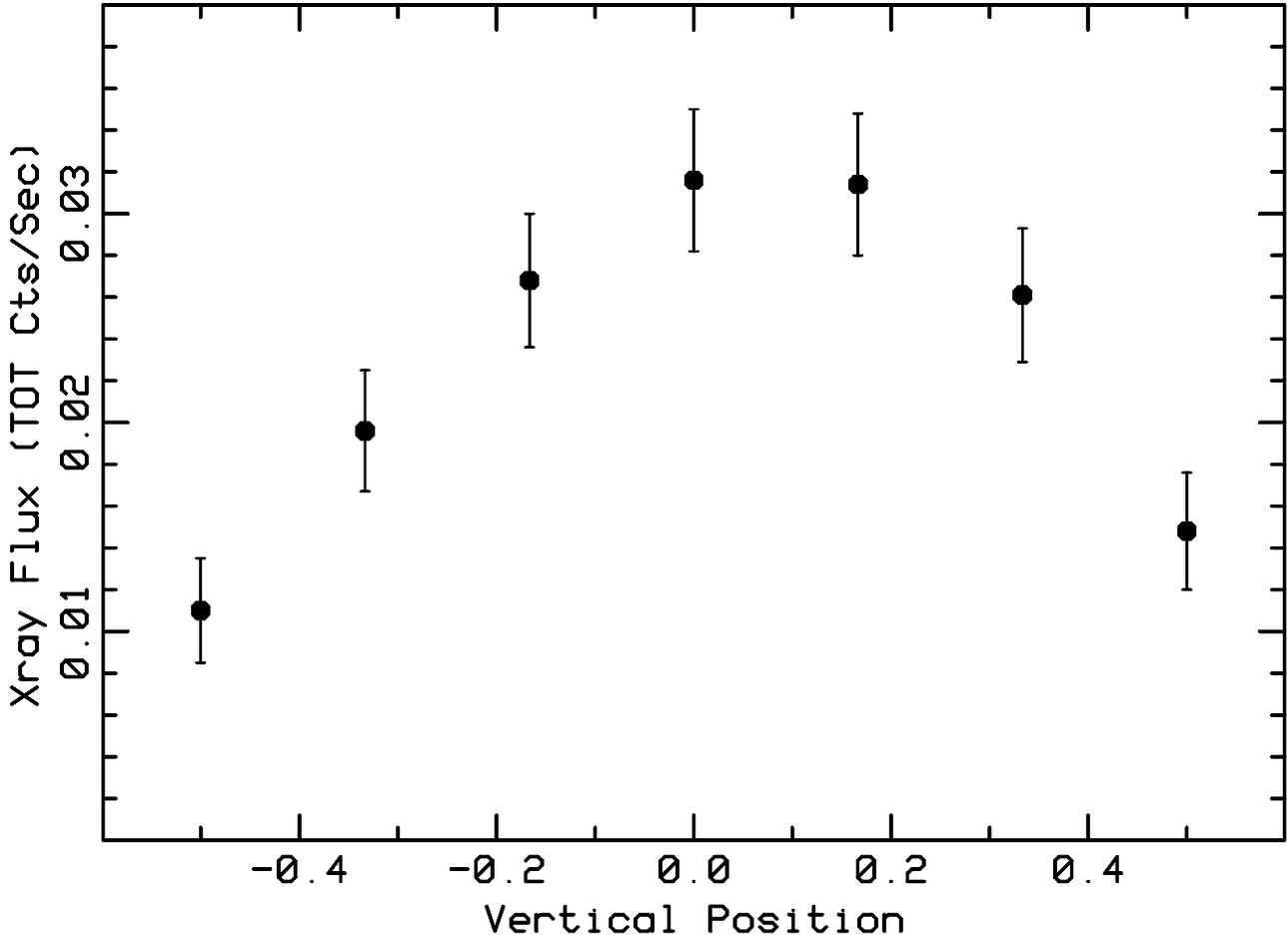}}}}
\caption[]
{
LSC emission fit to Map$\#2$ for pillbox models with central planes offset from 
the Galaxy.  Vertical position is expressed in terms of the thickness of the
pillbox.  The error bars are statistical only and are highly correlated.
}
\end{figure}

Figure 5 illustrates the dependence of results on the radial position of the 
Galaxy in the LSC.  Position is expressed in units of the supercluster
radius. The nominal value is $0.8$.  It is clear from the figure that the 
largest LSC emission is found for models with the Galaxy near the edge of the
LSC (note: negative values indicate models with 
centers in the direction opposite
Virgo).  The model with the nominal radial position has the smallest value of
$\chi^2$ while models with radial positions $\leq 0.5$ have significantly larger
$\chi^2$s.  For this test the results for Map$\#1$ differ 
somewhat in that models with radial positions $\geq 0.2$ have reasonable 
$\chi^2$s; however, the fit LSC emission was comparable for all these models.

\begin{figure}[ht]
\centerline{\rotate[r]{\vbox{\epsfysize=17cm\epsfbox{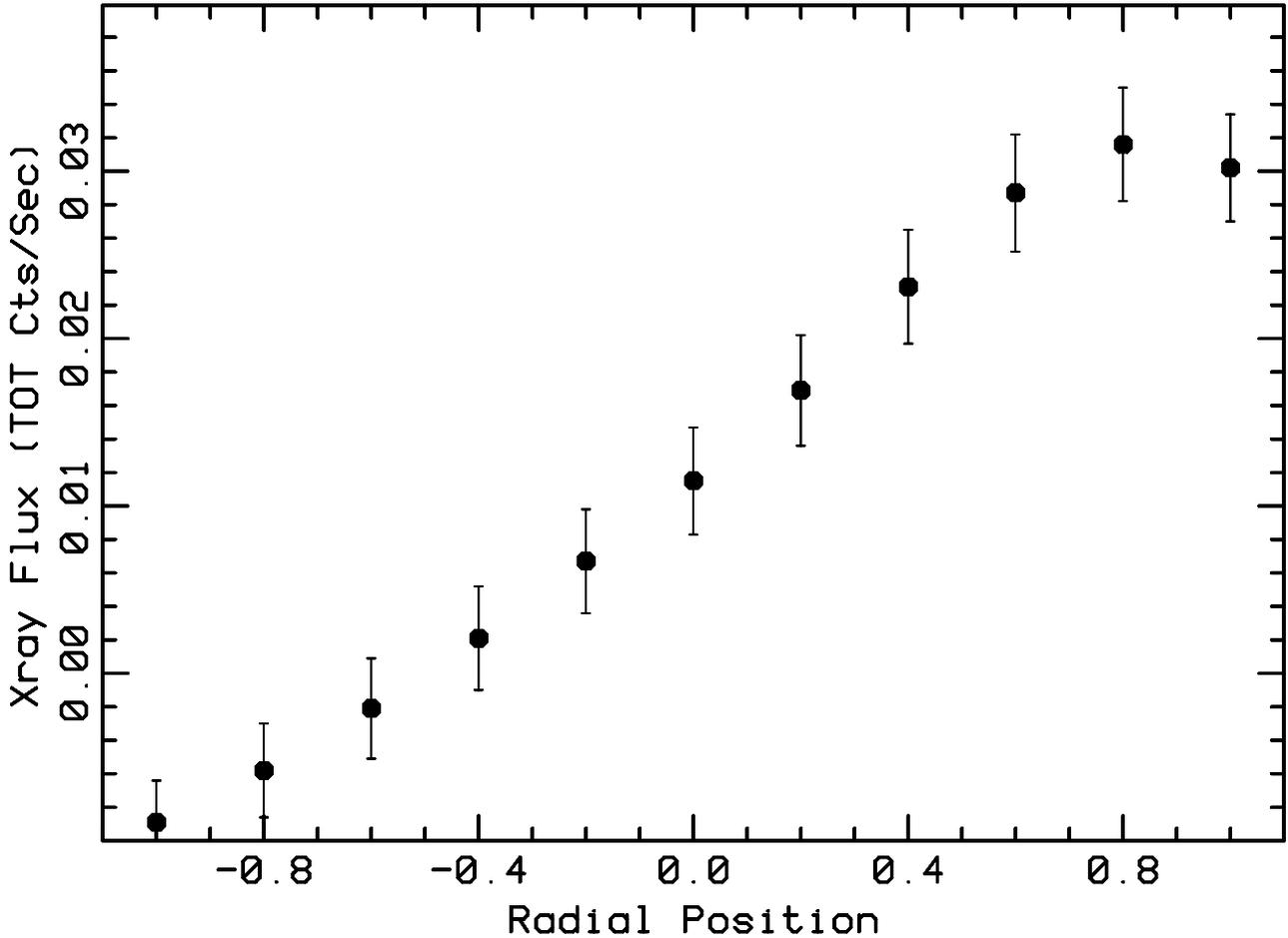}}}}
\caption[]
{
LSC emission fit to Map$\#2$ for pillbox models with centers displaced from the 
Galaxy.  Radial position is expressed in units of the supercluster
radius. The nominal value is $0.8$. Negative values indicate models with 
centers in the direction opposite Virgo.  The error bars are statistical only 
and are highly correlated.
}
\end{figure}

In order to test specifically whether or not a few isolated ``hot spots'' in
the supergalactic plane are responsible for the signal, the data is binned 
according to that emission predicted from the pillbox model and then plotted
in Figure 6 as a function of emission predicted from the model.  The scale of
the predicted emission is taken from Table 1 and the data is corrected for
all fit large-scale structure except for LSC emission.  If the signal is due to
hot spots in the direction of the center of the LSC, the extreme right hand
data point would fall well above the unity slope curve while other points
would fall well below the curve.  Within the limitations of
signal to noise, it appears that this is not the case.  Of course, the 
disposition of the data is sensitive to the binning.  While finer scale binning
does show a great deal of scatter, the linear trend is clear.
The data from Map$\#1$ shows the same trend with a bit more scatter.

\begin{figure}[ht]
\centerline{\rotate[r]{\vbox{\epsfysize=17cm\epsfbox{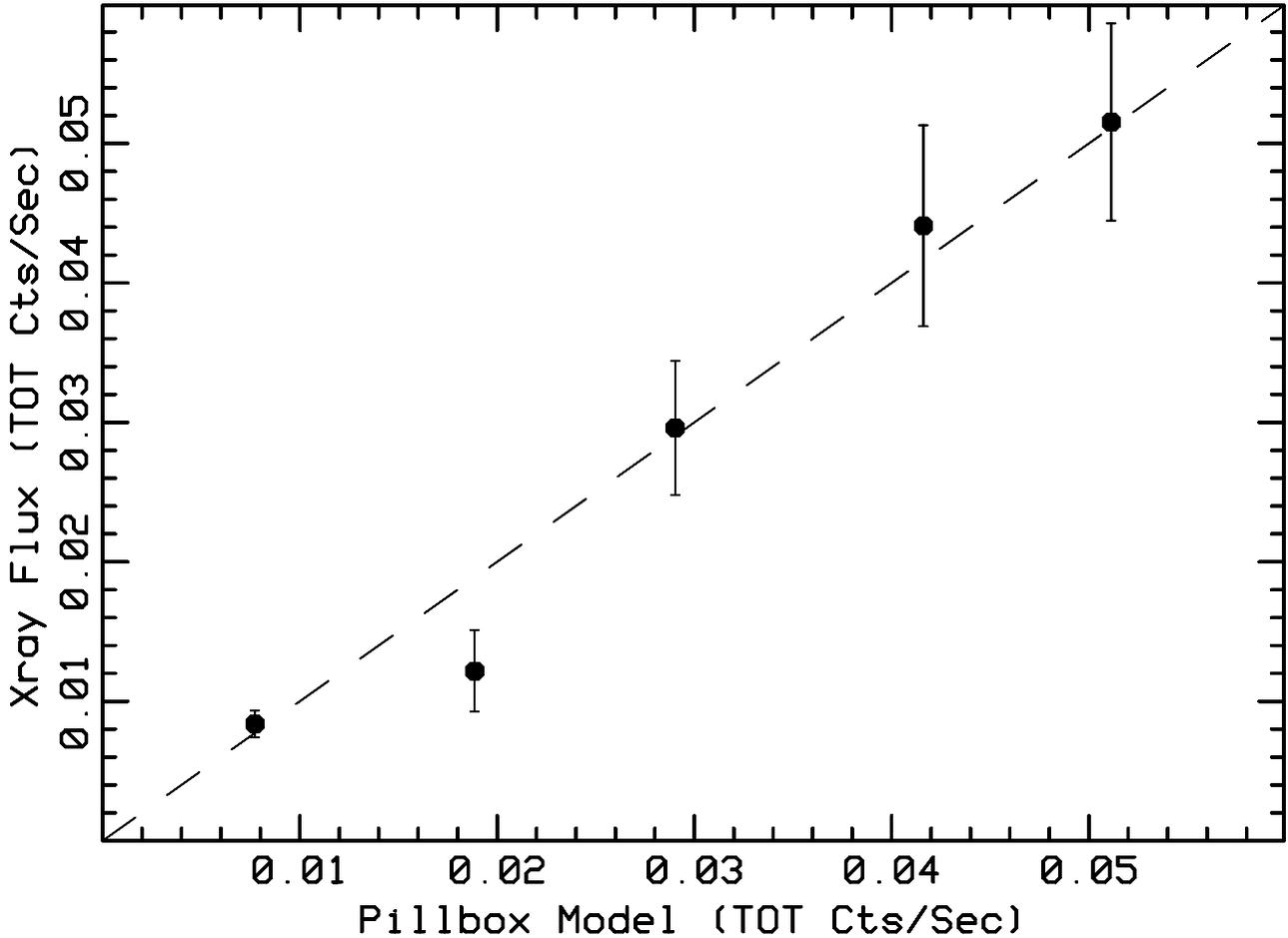}}}}
\caption[]
{
Average $2-10~keV$ flux of Map$\#2$ binned according to level of emission 
predicted from the pillbox model.  The predicted emission is taken from Table 1
and the map is corrected for all fit large-scale structure except for LSC 
emission.  The error bars are statistical only.
}
\end{figure}

Finally we investigate how sensitive the results are to varying the thickness
to diameter ratio of the pillbox model.  The answer is ``not very''.  For
Map$\#1$, only for ratios $H_{SC}/D_{SC}$ less than $1/12$ and greater than $1/4$
are the $\chi^2$s of the fits signficantly worse.  For Map$\#2$
these values are $1/16$ and $1/2$.  The values of LSC emission decrease
somewhat as $H_{SC}/D_{SC}$ increases; however, 
the statistical signficance of detection is roughly the same for all models
in this range.  One can only deduce a very rough value of the 
thickness to diameter ratio from the data and must rely on the optical structure
of the LSC (\cite{tul82}).  Clearly, the more important aspect of the models is
that the emission is enhanced in the general direction of Virgo.

\subsection{Windowing}
As discussed in \S2 pixels of the X-ray map were flagged if they were: 1) 
too close to the plane of the Galaxy; 2) if they were near a strong X-ray
source; or 3) if they were near a positive fluctuation in the XRB.  Failing
to remove the strong X-ray sources results in a very poor fit, 
$\chi_\nu^2 >30$; therefore, these cuts are absolutely necessary.  On the other
hand, the results of \S3.5 are relatively insensitive to the other two cuts.
The level of LCS emission changes by less than $30\%$ if the Galactic plane cut
is varied between $20^\circ$ and $50^\circ$ (the smallest fit value of
LSC emission is only $5\%$ less than that of Table 1).  Similarly,
the fit LSC emission does not change significantly from the minimally windowed 
Map$\#1$ to a map with only $20\%$ full sky coverage.

\subsection{X-ray Auto-Correlation Function}
As a check on how well the modelled structure matches the X-ray sky, the 
auto-correlation function (ACF) of the map was compared with that predicted by
the model.  One might expect that the unity values of the $\chi^2$'s of 
Table 1 indicate that the fit is quite good.  However, recall that this value 
was forced by assuming that all the excess variance was due to fluctuations
in the XRB.  If, on the other hand, the excess were due to unmodelled 
large-scale structure then the modelled ACF would not fit the data.  Figure 7 
shows that
the ACF is well modelled and that there appears to be little other large-scale
structure in the map.  The solid curve of Figure 2 was taken as the model of
the small-scale fluctuations in the XRB.  For $\theta > 10^\circ$ the data is 
binned in $13^\circ$ bins to reduce scatter.

\begin{figure}[ht]
\centerline{\rotate[r]{\vbox{\epsfysize=17cm\epsfbox{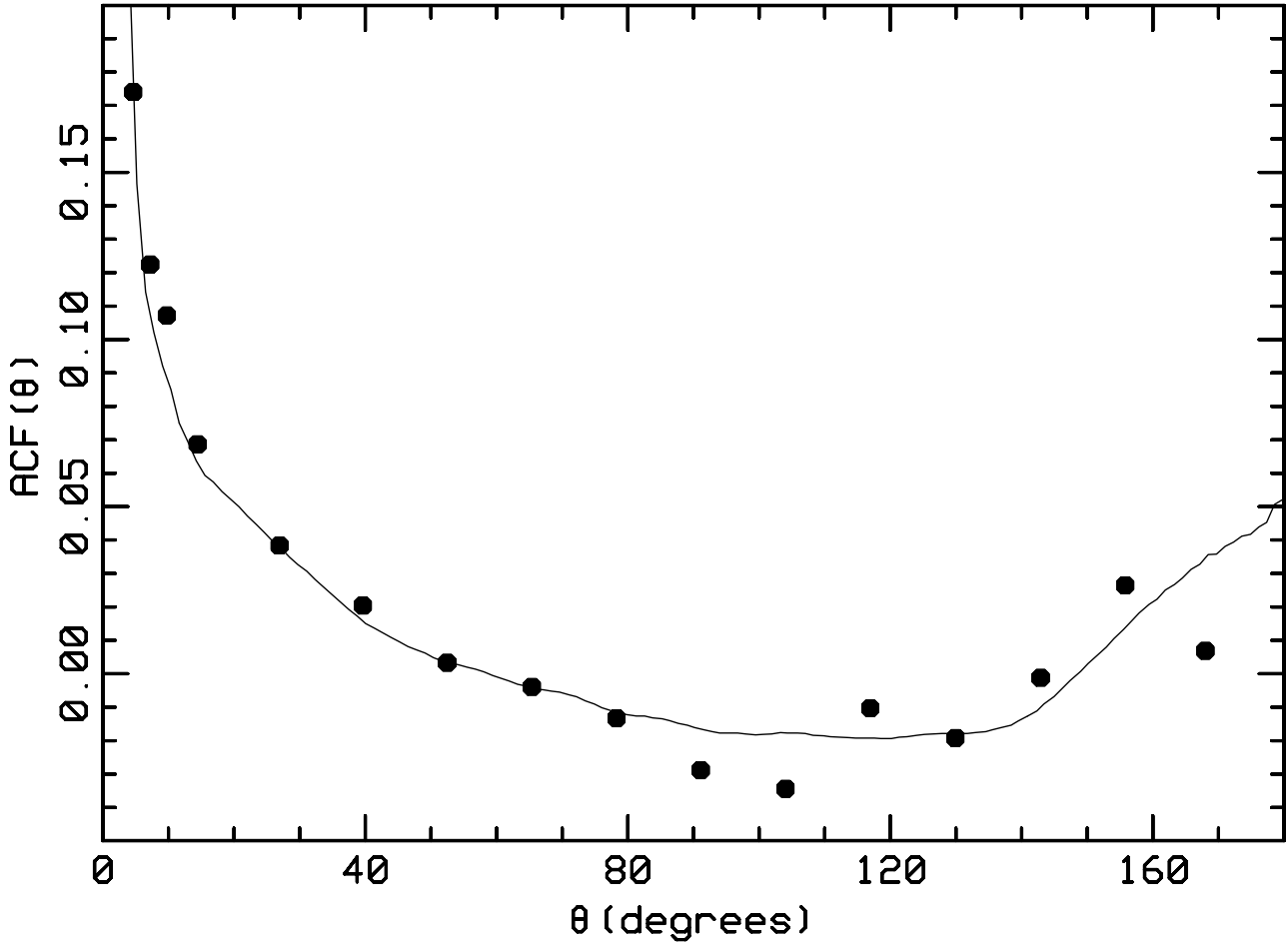}}}}
\caption[]
{
The autocorrelation function (normalized to unity) of Map$\#2$.  For $\theta
> 10^\circ$ the data is binned in $13^\circ$ bins to reduce scatter.  The solid
curve is the ACF for the model including all fit large-scale structure plus 
the fit of Figure 2 for the small-scale ($<10^\circ$) structure.
}
\end{figure}

\subsection{Point Sources}
Even if the above arguments indicate that there is X-ray emission associated 
with the plane of the LSC, it is not at all clear that this emission originates 
from diffuse gas.  After all, there are a great many galaxies and several 
clusters in this plane and all of these emit X-rays at some level.  However,
there are several indications that it is not the case that faint sources
contribute significantly to the X-ray emission in the LSC plane.
At a distance of $20~Mpc$ the cutoff flux of the present
map ($3\times 10^{-11}~ergs~s^{-1}~cm^{-2}$) corresponds to a luminosity of
$\sim 10^{42}~erg~s^{-1}$; therefore, only relatively weak (compared to AGN)
sources could contribute.  The observed upper limit of the 
average $2-10~keV$ emissivity of such sources in the local ($z \simle 0.1$) 
universe is $4 \times 10^{38}h~erg~s^{-1} Mpc^{-3}$ (\cite{miy94}) 
where $h$ is the Hubble constant in units of $100~km~s^{-1}~Mpc^{-1}$.  
On the other hand the pillbox model of the LSC and the amplitude of the X-ray 
flux in Table 1 imply an average X-ray emissivity of $3.0 \times 10^{39} ~ 
(R_{SC}/20~Mpc)^{-1} erg~ s^{-1} Mpc^{-3}$ (see \S6). Thus for a
Hubble constant of $60~km~s^{-1}~Mpc^{-1}$ the ratio of the observed 
LSC emissivity to the average emissivity of weak sources is $\simge 12$.  
For an LSC collapse factor is $\sim 10$, the observed emissivity could be due
to weak sources without seriously violating this constraint.

However, unified models of the XRB indicate that it is unlikely that weak
sources make a signficant contribution to the average X-ray emissivity.  For 
example, the model of Comastri et al. (1995) imply an average, local $2-10~keV$
emissivity of $7.6 \times 10^{38}h ~ erg~s^{-1} Mpc^{-3}$ which is consistent
with the observed value 
($8.6 \pm 2.4 \times 10^{38}h~erg~s^{-1} Mpc^{-3}$) derived
by Miyaji et al. (1994) from a cross-correlation analysis of the HEAO-1 A2 map
with galaxies from the IRAS survey.  The Comastri model (by design) accounts 
for the entire XRB.  According to this model, weak sources make a small
contribution to the local emissivity, i.e. 
$< 7.6 \times 10^{37}h~erg~s^{-1}~Mpc^{-3}$.  For $h=0.6$ this is a factor of
$\sim 65$ less than inferred for the LSC.  If weak sources account for the
LSC emissivity, then either Comastri et al. (1995) seriously underestimate
their numbers or the LSC collapse factor is much larger than observed for
visible galaxies.
While the above argument is suggestive, it is based on a unfied AGN model which
is by no means certain.  In fact, there is some spectulation that weak sources
do make a significant contribution to the XRB (\cite{yb98}; \cite{df97}).
Stronger constraints on the point source contribution to LSC emissivity come
from the consideration of two source catalogs, the Nearby Galaxies Catalog
(NBG) (\cite{tul88}) and the ROSAT All-Sky Survey Bright Source Catalog
(\cite{vog96}).

There are 2367 galaxies in the NBG catalog which consists of all galaxies with
known velocities $\le 3000~km~s^{-1}$.  
This catalog is dominated by two data sets: the Shapely-Ampes sample and the
all-sky survey of neutral hydrogen.  Although the catalog has severe 
incompleteness problems at velocities above $2000~km~s^{-1}$ the coverage is 
homogeneous across the unobscured part of the sky (\cite{tul88}).  In any case,
if point X-ray sources associated with galaxies in the LSC are an important
source of X-ray emissivity it seems reasonable that much of the emission
would be associated with the galaxies in this catalog.  To check this 
possibility, the X-ray map was further cleaned by removing $\sim 15$ square
degree regions (somewhat larger than the PSF of the map)
at the location of all of catalog galaxies.  Many of these regions lie in 
already windowed sections of the map.  After the additional cuts the sky 
coverage was reduced to $26\%$ for Map$\#1$ and $18\%$ for Map$\#2$.  
The reduced 
coverage results in an increase in noise; however, the levels of LSC 
emission are still marginally significant ($2~\sigma$ and $1.5~\sigma$ for
Map$\#1$ and Map$\#2$ respectively) and are somewhat larger than but 
consistent with the values in Table 1.

Another check of the level of contribution of galaxies from the NBG catalog
was made by fitting the LSC pillbox plus a monopole to the $15~deg^2$ regions 
surrounding 
the galaxies but with the appropriate windowing of Maps$\#1$ and $\#2$.  Before 
the fits the maps were corrected for a dipole, the Galaxy and instrumental 
drift.  The sky coverage for these fits was only $21\%$ and $15\%$; however, 
if these regions dominate LSC emission then the amplitude of the
fit should be considerably larger than in Table 1.  
This was not the case.  In fact, the fit 
amplitudes of LSC emission are essentially the same as those in Table 1.
Although the above tests are consistent with diffuse LSC emission, the Tully
galaxies (smoothed by the PSF) occupy a great deal of the LSC plane and so this
conclusion must be tempered. 

One final check on the galaxy contribution was to generate a map by assigning
an X-ray flux to each galaxy, convolve the resulting map with the PSF, and
then apply the windowing appropriate for Maps$\#1$ and $\#2$.  The X-ray flux of
the galaxies was obtained by cross-correlating the X-ray map with an
appropriately smoothed map of galaxy number density.  This was done in four 
different distance bins ($0-10;~10-20;~20-30;$ and $30-50~Mpc$) in order to
account for distance dependence of flux.  Only for the first bin was the
correlation statistically significant.  The X-ray emission from galaxies
was determined using the $1~\sigma$ upper limits of the cross-correlation 
for the appropriate distance bin.  The linear regression applied to
the resulting maps yields upper limits to LSC emission which are a factor of
3 below the levels of Table 1.  This result is another indication that
X-ray emission from galaxies is not the major component of X-ray emission from
the LSC plane.

A more direct indication of the contribution of point sources is the
ROSAT All-Sky Survey (RASS) Bright Source Catalog (\cite{vog96}).
There are 18,811 sources in this 
catalog down to a $0.1-2.4~keV$ flux limit of $0.05~cts/s$ 
($3 \times 10^{-13}~erg~s^{-1}~cm^{-2}$ for a photon spectral index of 
$\alpha = -2$ where $dN/dE \propto E^\alpha$).  At a brightness limit of 
$0.1~cts/s$ (8,547 sources) the catalogue represents a sky coverage of 92\%.
Two hardness ratios, $HR1$ and $HR2$, are defined by
$HR1 = (B-A)/(B+A)$ and $HR2 = (D-C)/(D+C)$ where $A$, $B$, $C$, and $D$ are
the count rates in passbands that correspond roughly to $0.1-0.4~keV$, 
$0.5-2.0~keV$, $0.5-0.9~keV$, and $0.9-2.0~keV$ (\cite{sno95}).
The average $HR2$ hardness ratio of the sources corresponds to
a photon spectral index of $\alpha \sim -2$; although, there are certainly 
sources that are much harder as well as sources that are much softer than 
this spectrum.  For $\alpha = -2$, the catalog limit corresponds to a 
$2-10~keV$ flux of $2 \times 10^{-13}~erg~s^{-1}~cm^{-2}$ which is about
two orders of magnitude below the Piccinotti et al. (1982) sources (see \S2).
If there is a signficant population of relatively
faint ($3 \times 10^{-13}$ to $3 \times 10^{-11}~erg~s^{-1}~cm^{-2}$) sources 
in the LSC, then a large fraction of these should show up in the Bright Source
Survey.  Even though the completeness level is not entirely uniform, the
catalog can still be used to give an indication of the level contamination 
by point sources.  There are too many RASS sources to simply window
them from the HEAO map ($3^\circ$ cuts around each source would window the
entire map); however, there are several ways of testing for the effects of 
RASS point sources.

To check the contamination due to relatively bright sources or clusters of 
sources we constructed a B-band ($0.5-2.0~keV$) map of RASS sources convolved
with the HEAO PSF.  If the resulting B-band flux in any pixel was greater than
$0.1~ct/sec$ (for $\alpha=-2$ this corresponds to a flux of 
$1 \times 10^{-12}~erg~s^{-1}~cm^{-2}$) then
a $15~deg^2$ region about the pixel was windowed from the two
previously windowed HEAO maps, i.e., Maps$\#1$ and $\#2$.
Note that a single source with a flux of $0.9~ct/sec$ is at the cutoff flux
after smoothing with the PSF.  The resulting heavily windowed maps, 
Map$\#1^\prime$ and  Map$\#2^\prime$, had sky coverages of $22\%$ and $17\%$ 
respectively.
When the linear regression of \S3.4 was performed on these two maps the fits to 
LSC emission were somewhat larger but not significantly different than the fits 
of Table 1.  The errors in the fits were about $50\%$ larger due to the reduced
sky coverage.  Similar results were obtained when the RASS cutoff flux level
was varied as well as for windowing individual sources (i.e. not convolved with
the PSF) with B-band fluxes above a cutoff level.  These results indicate that
the LSC emission indicated in Table 1 is not due to the brighter of the RASS
sources.

As another check on contaimination, every source in the RASS catalog was
assigned a $2-10~keV$ flux from its B-band flux by assuming a spectral 
index of $-1 < \alpha < -3$ which was deduced from $HR2$ hardness
ratio.  For faint sources the quantity HR2 is quite noisy and the computed
value of $\alpha$ might be either larger than -1 or smaller than -3.  For these
sources $\alpha$ was simply forced to be -1 or -3 which we consider to
the limits of the range for nearly all X-ray sources.  The B-band fluxes were
corrected for Galactic extinction using HI column density maps
(\cite{sta92}; \cite{dl90}) and the absorption coefficients 
of Morrison and McCammon (1983).  These corrections made little 
difference and, in any case, are probably less important than internal 
extinction in the sources themselves.  The resulting $2-10~keV$ flux map was
then convolved with the HEAO PSF and subtracted from the HEAO maps, Map$\#1$
and Map$\#2$.  The linear regressions applied to these corrected maps implied
LSC emission somewhat larger than but not significantly different from the
values in Table 1.  

As an upper limit to the contamination by RASS sources we applied the above map
correction assuming that every source has a very hard spectrum, 
i.e. $\alpha = -1$.  The linear regression fit to Map$\#2$ was essentially 
unchanged while the LSC emission fit to Map$\#1$ was reduced by about $40\%$.
Assuming every source has the average spectral index, i.e. $\alpha = -2$, yields
results that are virtually indistinguishable from those of Table 1.

At the catalog limit, $0.05$ cts/sec, the coverage is probably less uniform than
at a limit of $0.1$ cts/sec; therefore, we repeated the above corrections for
the brightest 8547 RASS sources ($> 0.1$ cts/sec).  The results
were virtually unchanged.

Because of the effects of non-uniform coverage and imperfect knowledge of the
spectra of individual sources, the above source
corrections to the $2-10~keV$ flux are not particularly accurate.  On the other
hand, they do indicate that the emissivity of weak X-ray sources in the LSC is
about an order of magnitude less that those implied in Table 1 and we conclude
that those results are not contaminated by point sources. 

\subsection{Clusters of Galaxies and the Great Attractor}
It is well known that rich clusters of galaxies are strong X-ray emitters and,
therefore, are a potential source of contamination.  
Distant clusters are indistinguishible
from the point sources discussed above.  In addition, 30 bright, nearby
clusters including Virgo, Coma and Centarus are among the Picinnotti et al. 
(1982) sources and have already been windowed from the map.  However, extended 
emission associated with nearby less compact clusters of galaxies might be 
considered a possible  source of contamination.  

Ebeling et al. (1997) used the ROSAT Brightest Cluster sample to fit a 
Schecter luminosity function to clusters and obtained a $2-10~keV~~L^*$ of
$1.2 \times 10^{45}~erg~s^{-1}$ and a power law exponent of -1.51.  Using these
values, the implied total $2-10~keV$ 
emissivity is $\varepsilon_x = 1.3 \times 10^{38}
~erg~s^{-1}~Mpc^{-3}$ which is a factor of $25$ less than that
implied by the values of Table 1 (see \S6).  On the other hand, the luminosity
of the Virgo cluster is $\sim 1.4 \times 10^{43}~erg~s^{-1}$ and sources this
bright would be cut from the windowed data if they were within
$60~Mpc$.  Again using the fit of Ebeling et al. (1997), the total
emissivity of clusters less luminous than the Virgo cluster is 
$\varepsilon_x = 1.6 \times 10^{37}~erg~s^{-1}~cm^{-2}~Mpc^{-3}$ or a factor
of 200 less than the implied LSC emissivity.  Even if this flux is increased by
the collapse factor of 10, it is still considerably less than that inferred 
for the LSC (see \S6).

As a further check on possible contamination by extended emission from galaxy
clusters we extended the windowing of the X-ray map to include $15^\circ$ 
diameter regions around the Virgo, Centarus, Coma, Fornax, and Ursa Major 
clusters as well as around the most dense regions of the Hydra, Pavo, Perseus,
and Ophiuchus galaxy clouds (\cite{tul88}).  Finally the diffuse emission found
by Jahoda \& Mushozky (1989) was eliminated by windowing a $40^\circ$ diameter
region about the Great Attractor.  Fits of the LSC emission to these further
windowed maps were only $\sim 15\%$ less that those of Table 1 while the 
errors in the fit were, of course, somewhat larger.  Therefore, there is no
indication that the fits of Table 1 are contaminated by either the cores of
clusters of galaxies or by diffuse emission from the Great Attractor.  

The windowing of more extensive regions of galaxy clouds in the LSC would not be
possible without cutting most of the plane of the LSC and, in any case, it is
not clear one should distinguish such emission with diffuse emission in the LSC.
This point is essentially the same as asking whether or not LSC emission is 
smooth or somewhat clumpy and will be addressed in \S6 below; however, the
bottom line is that the signal to noise is simply not high enough to be able to 
give a clear answer.

\section{Sunyaev-Zel'dovich Effect in the LSC}

If the diffuse X-ray emission in the LSC is due to Bremsstrahlung from a hot,
ionized intergalactic medium then one might expect its signature to be 
imprinted upon the Cosmic Microwave Background (CMB) via inverse Compton
scattering, i.e. the Sunyaev-Zel'dovich (SZ) effect 
(\cite{sz80}).  If the gas were uniformly distributed, the
profile in the CMB would be the same as in the X-ray except with a negative 
amplitude for observations on the Rayleigh-Jeans side of the blackbody
spectrum.  The CMB temperature decrement for radiation passing through a gas 
cloud of uniform electron density $N_e$, thickness $L$, and temperature $T_e$
is given by (e.g. \cite{hog92})
\begin{equation}
(\delta T/T)_{CMB} \approx 4.0\times 10^{-6}~\beta~l~n_e~t_e
\end{equation}
where $l = L/10Mpc$, $n_e = N_e/10^{-5}cm^{-3}$, $t_e = kT_e/10~keV$,
$\beta = (x{e^x+1\over e^x-1}-4)$, and $x=h\nu/kT_{CMB}$.  For $L=20~Mpc$,
$N_e=2.5\times 10^{-6}$, and $kT_e=10~keV$, the Rayleigh-Jeans decrement 
in the CMB is $|\delta T/T| \sim 4 \times 10^{-6}$.  This value is
somewhat smaller than the $10^\circ$ scale fluctuations observed by the 
COBE DMR experiment, comparable to other large-scale structure 
in the microwave sky, e.g. high latitude Galactic emission
and the expected intrinsic CMB quadrupole, and only marginally larger than the
instrument noise in the DMR data (\cite{ben96}).  Never-the-less, we
performed the same type of analysis as for the X-ray background described in
\S3 above.

The four year 53~GHz DMR map was obtained from the COBE data archive in
$2.6^\circ \times 2.6^\circ$, ecliptic, quadrilaterized spherical cube 
projection format (\cite{ben96}).  
This map was deemed more appropriate than the 31~GHz or
90~GHz maps because of a combination of low instrument noise, low Galaxy 
background, and moderately large $\beta$ (see eq. 5-1).  An 11 parameter linear
regression similar to that of \S3.4 was performed on the map after flagging
all pixels within $20^\circ$ of the Galactic plane and within $30^\circ$ of the
Galactic center.  The fit parameters were a monopole, a dipole, a quadrupole,
a secant law Galaxy model, and the amplitude of the canonical pillbox model
of the LSC.  As with the HEAO X-ray map, the sky coverage for the DMR maps
and, hence, instrument noise are not uniform across the sky.  In this case the
instrument noise per pixel is larger than the intrinsic sky fluctuations so 
the contribution to $\chi^2$ (see eq. 3-2) of each pixel is weighted inversely 
with noise variance of the pixel.  However,
the results did not change significantly when the analysis was repeated with
equally weighted pixels.  The formal fit to the pillbox amplitude
(normalized to $1~R_{SC}$) is
\begin{equation}
\delta T_{CMB} = -17\pm 5 ~\mu K
\end{equation}
where the uncertainty is statistical only and assumes uncorrelated instrument
noise.

As mentioned above, this level is comparable to other systematic structure in
the map and, therefore, should by no means be considered as a $3\sigma$
detection.  As an estimate of the significance of the result, the analysis was
repeated for 5000 model pillboxes with a uniform distribution of orientations
in the sky.  As in \S4.4 models lying within $30^\circ$ of the Galactic and
Supergalactic planes were disregarded.  The amplitude of the SZ effect of 
eq. 5-2 is more negative than 82\% of the rotated models indicating a 
significance of $\sim 1~\sigma$.  On the other hand, 80\% of the
$\chi^2$s of the trial fits exceed that of the fit of eq. 5-2 and 9\% 
of the trials have more negative SZ fits and smaller $\chi^2$s than the LSC
model.  If one includes models that lie within $30^\circ$ of the plane of the 
LSC these results do not change significantly.  Nor are they changed for 
analyses in which the pixels are weighted equally.  

Because of the low level of the signal only a few checks for systematics were
made.  The method of \S4.5 was used to exclude the possibility that a single
``hot'' or ``cold'' source accounted for the signal.  The LMC, SMC, and Orion 
Nebula were explicity excluded with no significant change in the fit.  Finally
the three DMR maps (31, 53, \& 90~GHz) were combined according to the
prescriptions suggested by Hinshaw et al. (1996) to minimize the effect of the
Galaxy.  These combinations have larger effective noise and so result in larger
statistical errors for the fits of the pillbox amplitude.  The fit amplitudes
varied from $-5~\mu K$ to $-22~\mu K$.
While it is clear the SZ effect due to hot gas in the LSC is not significantly
detected, we note that the upper limit is consistent with the amount of hot,
diffuse gas required to account for the diffuse X-ray emission discussed in
\S6.  We do find it intriguing that the fits correspond
to a decrement in the CMB as predicted by the SZ effect.

\section{Discussion}

If one takes seriously the hypothesis that hot, diffuse gas in the LSC is 
responsible for the diffuse X-ray emission claimed in \S3, then the strength
of the emission can be used to constrain the density and temperature of the gas.
For simplicity assume a uniform, isothermal gas with electron temperature $T_e$
and electron density $N_e$.  Then the $2-10~keV$ X-ray intensity due to 
Bremsstrahlung is given by (e.g. \cite{rl79})
\begin{equation}
I_x=\varepsilon_xL/4\pi=
4.0\times 10^{-9}~n_{e}^2t_{e}^{1/2}l~erg~s^{-1}cm^{-2}sr^{-1}
\end{equation}
where $L$ is the thickness of the emitting region, $l=L/10~Mpc$, 
$n_e = N_e/10^{-5}cm^{-3}$, and $t_e = kT_e/10~keV$.  Primordial element
abundances are assumed; however, the coefficient in eq. 6-1 only increases 
by a factor of 1.14 for solar abundance.  From Table 1,
the amplitude of the LSC emission normalized to $1~R_{sc}$ is 
$\sim 3.3 \times 10^{-2}~TOT~cts/s/4.5~deg^2$ which corresponds for a $10~keV$ 
Bremsstrahlung spectrum to
\begin{equation}
I_x = 5.0 \times 10^{-10}~erg~s^{-1}~cm^{-2}~sr^{-1}.
\end{equation}
From eqs. 6-1 and 6-2 (with $L=R_{SC}$)
\begin{equation}
N_e = 2.5 \times 10^{-6}~(R_{SC}/20~Mpc)^{-1/2}(kT_e/10~keV)^{-1/4}~cm^{-3}.
\end{equation}
The implied gas density is only weakly dependent on $R_{SC}$ and $T_e$.
Moreover, it is reasonable to assume that $R_{SC} \sim 20~Mpc$ (roughly the
distance to the Virgo cluster) and that $kT_e \sim 10~keV$.  A temperature
much greater than 10 keV would exceed the virial temperature of the LSC while
a temperature much less than 10 keV would have rendered the gas undetectable by
HEAO.  As a rough consistency check, the X-ray data were split into ``soft''
(2-5 keV) and ``hard'' (5-10 keV) components (\cite{ajw94}) and the 
``pillbox'' fits of \S3 repeated on the subdivided data sets.  The ratio of the
fit amplitudes (in cts/s) in these two bands is 
${\dot N_{soft}}/{\dot N_{hard}} \simle 1$.  Although the uncertainty in this 
ratio is considerable, it is consistent with an electron temperature of 
$\simge$ 10 keV but inconsistent with temperatures $< 3 keV$.  Therefore, it 
seems unlikely that the implied electron number density could be much 
different than $N_e \sim 2$ to $3 \times 10^{-6}$.  It is interesting to note 
that this value is roughly an order of magnitude larger than the mean number
density of baryons in the universe and is consistent with a collapse factor
of 10 which is roughly the aspect ratio of the LSC.

If this hot gas were distributed uniformly within a 40 Mpc diameter by 5 Mpc
thickness supercluster, the implied total mass is 
$\sim 4 \times 10^{14}~ M_{\odot}$.  From dynamical considerations, 
Shaya, Peebles, \& Tully (1995) have estimated the total mass within a distance 
of 40 Mpc (about 40 times the volume of the LSC) is about $7 \times 10^{15}~M_
{\odot}$.  On the other hand, total mass in stars in the same volume is about
$1 \times 10^{14}~M_{\odot}$, which is an order of magnitude too small to
account for the baryonic matter which should be present.  While the mass of the 
hypothetical hot gas in the LSC is insufficient to account for the dynamical 
mass in the local universe it may well make up the bulk of the baryonic matter.

These estimates rely on the gas being uniformly distributed.  Since 
Bremsstrahlung is proportional to $N_{e}^2$, the emission is enhanced if there 
is any clumping of the gas.  For example, if the gas is contained in 50\% of 
the volume of the disk of the LSC, i.e. a clumping factor of 2, then the 
implied mean density decreases by a factor of $1/\sqrt 2$.

As discussed in \S5 above, the presence of hot, ionized gas is imprinted via
the SZ effect on the cosmic microwave background.  For 10 keV gas with a density
of $2.5 \times 10^{-6}~cm^{-3}$ and a thickness of 20 Mpc, the expected 
decrement in the CMB is $\sim -11~\mu K$ for a Rayleigh-Jeans spectrum.  While
the fit of a pillbox LSC to the 53 GHz COBE map is consistent with this 
prediction, the systematic structure in the map is large and the agreement 
should be considered at best a $1~\sigma$ confirmation.  

If superclusters (SCs) with hot gas are common in the universe then their 
combined SZ effects would result in fluctuations in the CMB which might be 
confused with intrinsic CMB fluctuations (\cite{hog92}).  This has been 
demonstrated not to be the case for the COBE results (\cite{bj93}; 
\cite{ben93}).  To see how many SCs are allowed by this constraint 
suppose that a fraction $f$ of all bayonic matter is contained in SCs with 
diameters of 40 Mpc and thicknesses 5 Mpc.  Then the number density of
SCs is $n_{SC} = f/bV_{SC}$ where $V_{SC}$ is the volume of an SC and 
b is the collapse factor.  If each results in a temperature decrement of
$\delta T_{CMB} \sim 2$ to $3~\mu K$ (for a path length equal to the thickness 
of the SC) then the $rms$ fluctuations of a distribution of
SCs should be $\sim 2.5\times \sqrt{N_{SC}}~\mu K$ where 
$N_{SC} \approx \pi R_{sc}^2 n_{SC}r$
is the number of SCs along the line of sight out to a distance $r$.  
To compare with the COBE DMR data
for which $\delta T_{rms} \sim 35~\mu K$ we set $r \sim 400 Mpc$ at which a
40 Mpc SC would subtend an angle about equal to the COBE beam size.  
Even if $f=1$, i.e. all the baryons
in the universe are in the form of hot, diffuse gas in SCs, the fluctuations
caused by the SZ effect would be $\sim 7~\mu K$ or about 1/5 the level of
the CMB fluctuations.  In all likelihood, the fraction $f$ is much smaller than
unity.  From the supercluster (SC) catalog of Batuski and Burns
(1985), Rephaeli (1993) estimated the local density of SCs to be 
$5\times 10^{-8}~h^3~Mpc^{-3}$.  If these SC's have volumes and densities of
that inferred for the LSC, then only about $0.001$ of the baryonic matter in
the universe is contained in SCs, i.e. $f=0.001$.  Then the SZ fluctuations
would be quite small, $\simle 1~\mu K$.  In any case, the spectrum of SZ 
fluctuations differs significantly from those intrinsic to the CMB.

Another way to detect the presence of the SZ effect is via a spectral
distortion of the CMB which is quantified by the Compton $y$ parameter, 
$y = \int (kT_e/m_ec^2)N_e\sigma_Tdl$ where $T_e$ is electron temperature,
$m_e$ is the electron mass, $N_e$ is electron density, $\sigma_T$ is
the Thompson scattering cross-section, and $l$ is path length 
(e.g. \cite{rep93}).  Assuming an isothermal gas this becomes
$y = (kT_e/m_ec^2)\sigma_T\int N_edl$.  A rough approximation of the integral is
$f$ times the mean baryon density times the Hubble radius, i.e.,
$fn_bcH_0$.  Then for $kT_e = 10~keV$, $y \sim 4 \times 10^{-5}~f$.  The limit 
on $y$ from the COBE data is (\cite{fix96}) $y \simle 10^{-5}$ which 
implies that $f \simle 1/4$.  If $f \sim 0.001$ as inferred from the local 
density of SCs, then the presence of hot gas in these SCs is consistent with 
the upper limit to the spectral distortion of the CMB. 

\section{Conclusions}

Evidence is presented in this paper for X-ray emission associated with the 
plane of the local supercluster (LSC).  While this has been suggested 
previously (\cite{jah93}), we argue that the emission is unlikely to be
produced by individual sources but rather is diffuse in nature.  This implies
that there is a great deal of hot ($\sim 10~keV$), diffuse 
($\sim 2.5 \times 10^{-6}~cm^{-3}$) gas in the LSC 
and that the gas may account for the
bulk of baryonic matter in the local universe.  The presence of such gas would
be imprinted on the cosmic microwave background (CMB) as a Sunyaev-Zel'dovich 
temperature decrement of $\sim -10~\mu K$.  While the COBE 53 GHz map is
consistent with such structure, other systematics preclude the positive
identification of this component.  Even if superclusters are relatively 
plentiful in the universe and hot gas in them is common, the resulting 
fluctuations in the CMB would be small relative to the those found in the 
COBE data and, therefore, unlikely to compromise the cosmological implications
of those fluctuations.

It should be emphasized that the existence of hot, diffuse gas in the LSC is by
no means firmly established.  The results presented here constitute only a
$2$ to $3~\sigma$ effect and $3~\sigma$ results have a history of disappearing. 
Also the source of the X-ray emission may not be diffuse, hot gas; although, 
we argue that point sources probably do not
account for it.  None-the-less, the results are tantalizing and are consistent
with the density and temperature of gas that might be expected to inhabit the
intergalactic medium.  It is unlikely that more detailed analyses of the HEAO
and COBE data will shed more light on the situation.  The signals to noise of 
these maps are simply not good enough.  However, the next generation of 
X-ray satellites with higher angular resolution, better frequency resolution,
and higher sensitivities will likely be able to either confirm or refute the 
suggestions made in this paper as well as be able to detect diffuse emission
in other relatively nearby superclusters.  Finally the new CMB satellites
scheduled for launch in the next few years (i.e. MAP and PLANCK) should have the
angular and frequency resolution required to distinguish an SZ effect in the
LSC from intrinsic CMB fluctuations if the level of the effect is that 
suggested by this paper.

\acknowledgements
I would like to acknowledge Keith Jahoda who is responsible for constructing the
the HEAO I A2 X-ray map and who provided me with several data handling programs.
I would also like to acknowledge Ruth Daly who initially suggested this project.
Much of this work was completed at the Princeton University Gravitation and 
Cosmology computing cluster where I benefitted greatly from the subroutines of
Ed Groth.  Finally I would like to acknowledge the hospitality afforded me at
the Institute for Advanced Study where I collected many of my thoughts for this
paper.  This work was supported in part by NASA grant NAG 5-3015, the Monell
Foundation, and (through Princeton University) NSF grant PHY-9222952.

\end{document}